\documentclass[journal,twoside,web]{ieeecolor}

\newcommand{\eqal}[2]{\\\begin{equation}\begin{aligned} \label{eq:#1} #2 \end{aligned}\end{equation}\\}
\newcommand{\Tone}{T\textsubscript{1}}
\newcommand{\Ttwo}{T\textsubscript{2}}

\newcommand{\ToneTtwo}{T\textsubscript{1}~$\rightarrow$~T\textsubscript{2}}

\newcommand{\TtwoTone}{T\textsubscript{2}~$\rightarrow$~T\textsubscript{1}}
\newcommand{\TtwoPD}{T\textsubscript{2}~$\rightarrow$~PD}

\newcommand{\PDTtwo}{PD~$\rightarrow$~T\textsubscript{2}}

\usepackage{subcaption}
\usepackage{multirow}
\usepackage{tmi}
\usepackage{cite}
\usepackage{amsmath,amssymb,amsfonts}
\usepackage{algorithmic}
\usepackage{graphicx}
\usepackage{textcomp}
\usepackage{amsfonts}
\usepackage[figurename=Fig.]{caption}

\def\BibTeX{{\rm B\kern-.05em{\sc i\kern-.025em b}\kern-.08em
    T\kern-.1667em\lower.7ex\hbox{E}\kern-.125emX}}
\begin{document}
\title{Semi-Supervised~Learning~of Mutually~Accelerated~MRI~Synthesis without~Fully-Sampled~Ground~Truths}
\author{Mahmut Yurt, Salman Ul Hassan Dar, Muzaffer Özbey, Berk Tınaz, Kader Karlı Oğuz, Tolga Çukur
\thanks{This study was supported in part by a TUBITAK 1001 Research Grant (118E256), a TUBA GEBIP 2015 fellowship, and a BAGEP 2017 fellowship (Corresponding author: Tolga Çukur).}
\thanks{M. Yurt, S. U. H. Dar, M. Özbey, and T. Çukur are with the Department of Electrical and Electronics Engineering, Bilkent University, Ankara, Turkey (e-mails: \{mahmut, salman, muzaffer, cukur\}@ee.bilkent.edu.tr).}
\thanks{B. Tınaz is with the Department of Electrical and Computer Engineering, University of Southern California, Los Angeles, California, USA (email: tinaz@usc.edu).}
\thanks{K. K. Oğuz is with the Department of Radiology, Hacettepe University, Ankara, Turkey (email: kkarlioguz@gmail.com).}}

\maketitle

\begin{abstract}
Learning-based synthetic multi-contrast MRI commonly involves deep models trained using high-quality images of source and target contrasts, regardless of whether source and target domain samples are paired or unpaired. This results in undesirable reliance on fully-sampled acquisitions of all MRI contrasts, which might prove impractical due to limitations on scan costs and time. Here, we propose a novel semi-supervised deep generative model that instead learns to recover high-quality target images directly from accelerated acquisitions of source and target contrasts. To achieve this, the proposed model introduces novel multi-coil tensor losses in image, k-space and adversarial domains. These selective losses are based only on acquired k-space samples, and randomized sampling masks are used across subjects to capture relationships among acquired and non-acquired k-space regions. Comprehensive experiments on multi-contrast neuroimaging datasets demonstrate that our semi-supervised approach yields equivalent performance to gold-standard fully-supervised models, while outperforming a cascaded approach that learns to synthesize based on reconstructions of undersampled data. Therefore, the proposed approach holds great promise to improve the feasibility and utility of accelerated MRI acquisitions mutually undersampled across both contrast sets and k-space.  
\end{abstract}


\section{Introduction}
\label{sec:introduction}
\renewcommand{\thetable}{\arabic{table}}   
MRI is a clinical powerhouse in neuroimaging due to its noninvasiveness and excellent soft-tissue contrast. Its unique ability to image the same anatomy under a diverse set of tissue contrasts empowers it to accumulate complementary diagnostic information within a single exam session \cite{bastiaan2009,bakas2017}. However, prolonged scans and increased costs associated with multi-contrast protocols often limit the diversity and quality of MRI exams \cite{thukral2015,krupa2015}. A promising solution against this limitation is synthesis of missing or unacceptably low-quality images within the protocol from available high-quality images \cite{iglesias2013}. Multi-contrast MRI synthesis methods can enhance radiological assessments as well as image analysis tasks such as registration, segmentation, or detection \cite{lee2020,huo2018,roy2013}.  

In recent years, there has been emerging interest in learning-based MRI synthesis based on deep neural networks, given their state-of-the-art performance in other computer vision \cite{goodfellow2014,mirza2014,zhu2017,isola2017,choi2017} and medical imaging tasks \cite{shin2016,schlemper2018,gupta2018}. An earlier group of studies proposed deep models with convolutional neural networks (CNNs) to learn nonlinear latent representations that mediate conversion from source to target images \cite{chartsias2017,sevetlidis2016,joyce2017,wei2019,bowles2016}. These studies typically involved encoder-decoder architectures, where the encoder embeds hierarchical image features onto a latent space that is later used by the decoder to recover the target image \cite{chartsias2017,sevetlidis2016,joyce2017,wei2019,bowles2016}. For improved capture of structural details, a second group has proposed deep architectures based on conditional generative adversarial networks (GAN) \cite{dar2019,sharma2019,yurt2021,zhou2020,nie2018,lee2019colla,li2019,yu2019,dar2020prior,armanious2018,beers2018,hagiwara2019,yurt2020provo,wang2020,yu2018}, where the generator that performs the source-to-target mapping benefits from the game-theoretic interplay with the discriminator \cite{goodfellow2014}. Pioneering studies have exploited pixel- or feature-wise correspondence between source-target images in an adversarial setup \cite{dar2019,hagiwara2019,beers2018}. Later studies have proposed unified models capable of multiple types of contrast conversion \cite{sharma2019,lee2019colla,li2019}, or multi-tasking frameworks \cite{wang2020,lee2019colla,sharma2019} to reduce computational complexity. These previous studies have collectively highlighted the immense potential of learning-based synthesis in multi-contrast MRI. That said, both CNN and GAN models are canonically trained in a fully-supervised setup based on pixel-wise, adversarial or perceptual losses between synthesized and ground truth target images. Supervised models require large datasets of high-quality images from Nyquist-sampled source and target acquisitions, paired within subjects \cite{chartsias2017,sharma2019,dar2019}. Yet, compilation of paired, high-quality datasets might prove impractical due to scan time and cost considerations \cite{thukral2015,krupa2015}. As such, there is a dire need for methods with lower reliance on supervision to improve practicality of learning-based MRI synthesis. 

Recent efforts to lower supervision requirements in MRI synthesis have predominantly focused on model training in the absence of paired images across subjects. For unpaired training, a successful approach has been to replace pixel-wise losses in GAN models with cycle-consistency, shape-consistency or mutual information losses \cite{olut2018,lee2019colla,sohail2019,dar2019,wolterink2017,ge2019}. Similar to supervised models, unpaired models that unify multiple contrast conversion tasks have also been introduced to reduce computational complexity \cite{sohail2019,choi2017}. As an alternative, \cite{jin2018,wang2020semi} have proposed a hybrid method where the model is trained on a composite dataset with both paired and unpaired samples. These previous methods have increased  the capacity of synthesis models to learn from unpaired data, but they still leverage high-quality MR images reconstructed from fully-sampled k-space acquisitions. While training of MRI reconstruction models from undersampled data has received recent interest \cite{yaman2020,cole2020unsupervised}, to the best of our knowledge, no prior study has considered learning of MRI synthesis models from undersampled source or target acquisitions. 

Here, we propose a novel semi-supervised deep generative model for multi-contrast MRI synthesis, namely ssGAN, to avoid reliance on fully-sampled k-space acquisitions. The proposed model is trained directly on undersampled acquisitions, and it generates high-quality target images given undersampled multi-coil source acquisitions. To do this, ssGAN introduces novel multi-coil tensor losses in image, k-space and adversarial domains. These selective losses are based only on acquired k-space samples, and randomized sampling masks are used across subjects to capture relationships among acquired and non-acquired k-space regions. Comprehensive experiments performed on brain MRI clearly demonstrate that ssGAN achieves equivalent performance to gold-standard models based on fully-supervised training across a broad range of acceleration factors. Meanwhile, ssGAN outperforms a cascade-model that first reconstructs undersampled acquisitions using compressive sensing \cite{SparseMRI,CSMRI,spirit}, and then trains a learning-based synthesis model.  
\subsubsection*{Contributions}
\begin{itemize}
\item To the best of our knowledge, this is the first semi-supervised learning method for multi-contrast MRI synthesis that performs model training based on undersampled source and target acquisitions.
\item The proposed method synthesizes target images directly from undersampled multi-coil source acquisitions. 
\item The proposed method introduces novel multi-coil tensor losses in image, k-space and adversarial domains, selectively expressed based on acquired k-space samples in target contrast acquisitions. 
\item The proposed method substantially lowers data requirements in MRI synthesis by enabling model training and inference from undersampled acquisitions. 
\end{itemize}

\begin{figure*}
\centerline{\includegraphics[width=0.73\textwidth]{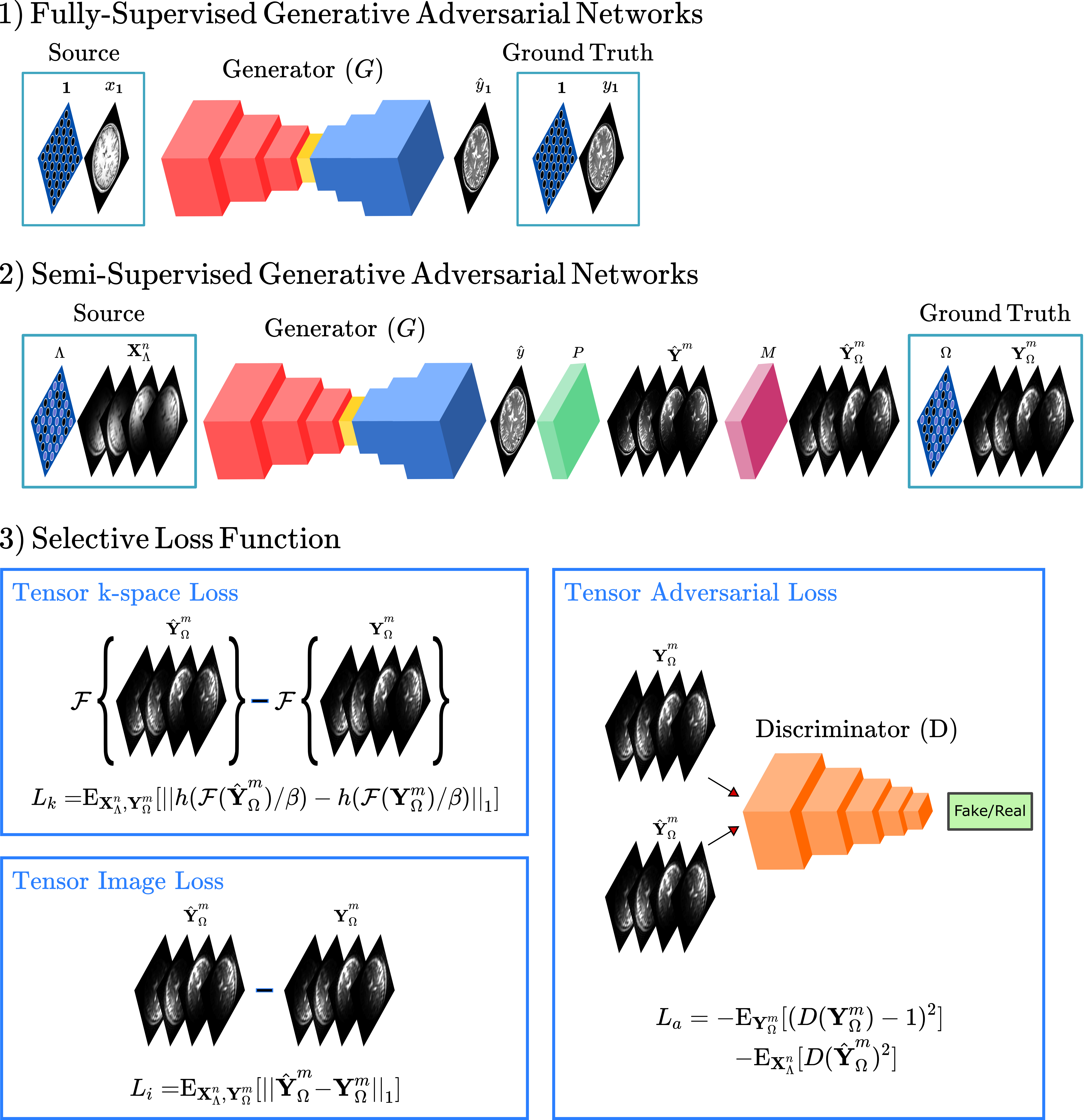}}
\caption{Illustration of the proposed semi-supervised ssGAN model. As opposed to fully-supervised models that demand Nyquist-sampled acquisitions for training (1), ssGAN learns to synthesize high-quality images given a dataset of undersampled source and target acquisitions (2). ssGAN initially synthesizes a coil-combined target image that is backprojected onto individual coils via sensitivity maps. These multi-coil target images are subsampled in Fourier domain with the target acquisition mask in order to define the selective multi-coil tensor losses in image, k-space and adversarial domains (3).}
\label{fig:mainfig}
\end{figure*}

\section{Methods}
In this section, we first overview basics of generative adversarial networks, and the foundation of the proposed architecture for semi-supervised multi-contrast MRI synthesis. We then describe in detail the datasets and experiments conducted to evaluate the proposed methodology.  
\subsection{Generative Adversarial Networks}
Generative adversarial networks (GANs) \cite{goodfellow2014} are deep generative models comprising a pair of competing subnetworks: a generator ($G$) and a discriminator ($D$). $G$ aims to map a random noise vector $z$ to a sample resembling a target domain distribution, whereas $D$ aims to distinguish between real and fake samples of the target domain \cite{goodfellow2014}. These two subnetworks are alternately trained via an adversarial loss function, formulated as follows:
\eqal{}{
L_{GAN}=-\mathrm{E}_y[(D(y)-1)^2]-E_z[D(G(z))^2]
} 
where $\mathrm{E}$ denotes expectation, and $y$ is an arbitrary real sample in the target domain. Upon convergence, $G$ is expected to generate realistic target domain samples that $D$ cannot tell apart from the real ones \cite{goodfellow2014}. While the initial GAN models generated target samples from a random noise vector, later studies have demonstrated success in image-to-image translation with conditional GAN (cGAN) models that additionally receive as input a source domain image $x$ \cite{mirza2014}. The adversarial loss function is therefore modified by conditioning $G$ on $x$: 
\eqal{}{
L_{cGAN}=-\mathrm{E}_{x,y}[(D(y)-1)^2]-E_x[D(G(x))^2]
} 
When spatially aligned source-target images are available, a pixel-wise loss can be further included \cite{isola2017}:
\eqal{}{
L_{cGAN}=&-\mathrm{E}_{x,y}[(D(y)-1)^2]-E_x[D(G(x))^2]\\&+\mathrm{E}_{x,y}[||y-G(x)||_1]
}
Several studies have demonstrated variants of cGAN models on multi-contrast MRI that synthesize target contrast images from source contrast images of the same underlying anatomy \cite{dar2019,sharma2019,zhou2020,nie2018,lee2019colla,li2019,yu2019,armanious2018,beers2018,hagiwara2019,wang2020,yu2018}. These models typically learn the source-to-target mapping in a fully-supervised setup. A comprehensive training set is needed containing high-quality source and target images reconstructed from fully-sampled k-space acquisitions ($x_\mathbf{1}$, $y_\mathbf{1}$), where $x_{\mathbf{1}}$ is an arbitrary source, $y_{\mathbf{1}}$ is an arbitrary target image in the training set, and $\mathbf{1}$ denotes the the sampling mask for Nyquist-sampled acquisitions. These fully-supervised models have demonstrated state-of-the-art performance for synthetic multi-contrast MRI. However, they are limited due to reliance on fully-sampled acquisitions that might prove impractical. Therefore, there is a critical need for methods that can directly learn from undersampled MRI data.  
\subsection{Semi-Supervised Generative Adversarial Networks}
Here, we propose a novel semi-supervised GAN model, namely ssGAN, to mitigate the dependency of MRI synthesis models on supervised training with Nyquist-sampled source and target acquisitions. ssGAN is trained on undersampled acquisitions of source and target contrasts, and it synthesizes multi-coil target images directly from undersampled multi-coil acquisitions of the source contrast. To do this, ssGAN introduces novel selective loss functions expressed based on only the acquired subset of k-space samples in the target contrast (Fig. \ref{fig:mainfig}). Details regarding the optimization objectives of ssGAN are provided in the remainder of this section. 

ssGAN receives as input Fourier reconstructions of either fully-sampled or undersampled acquisitions of the source contrast, and learns to synthesize high-quality images of the target contrast. The generator $G$ in ssGAN produces target contrast images via a forward mapping: 
\eqal{}{
&\,\,\,\,\,G(\textbf{X}^n_\Lambda)=\hat{y}, \,\,\,\, \mathrm{with}\,\,\,\,\textbf{X}^n_\Lambda&=\{x^1_\Lambda,\dots,x^n_\Lambda\}
}
where $\textbf{X}^n_\Lambda$ denotes multi-coil source contrast images acquired with a k-space sampling mask $\Lambda$, $n$ denotes the number of receive coils with sensitivity maps $\hat{\textbf{C}}_\textbf{X}^n$ computed via ESPIRiT \cite{espirit}, and $\hat{y}$ denotes the synthesized coil-combined target contrast image. Note that ssGAN considers that only undersampled acquisitions of the target contrast are available, where $\textbf{Y}^m_\Omega=\{y^1_\Omega,\dots,y^m_\Omega\}$ denotes Fourier reconstructions of multi-coil target acquisitions collected with a sampling mask $\Omega$ and $m$ receive coils of true coil sensitivities $\textbf{C}_\textbf{Y}^m$. As no high-quality reference for the target contrast image is assumed, ssGAN expresses novel selective loss functions based on only the acquired subset of k-space samples. To do this, the synthesized coil-combined image is first projected onto individual coils as follows: 
\eqal{}{
\hat{\textbf{Y}}^m=P(\hat{y},\hat{\textbf{C}}_\textbf{Y}^m)=\hat{y}\cdot\hat{\textbf{C}}_\textbf{Y}^m
}
where $\hat{\textbf{Y}}^m$ denotes the synthesized multi-coil target contrast images, $\hat{\textbf{C}}^m_\textbf{Y}$ denotes estimated coil sensitivity maps computed via ESPIRiT \cite{espirit}, and $P$ is the operator that performs the coil projection in the image domain as dot product takes vectors and outputs a scalar, element-wise multiplication between the input image and coil sensitivity maps. The multi-coil target image projections are then subjected to the binary sampling mask in Fourier domain: 
\eqal{}{
\hat{k}_{Y^m_\Omega}&=M(\mathcal{F}(\hat{\textbf{Y}}^m),\Omega)=\mathcal{F}(\hat{\textbf{Y}}^m)\cdot\Omega\\
\hat{\textbf{Y}}^m_\Omega&= \mathcal{F}^{-1}(\hat{k}_{Y^m_\Omega})
}
where $\mathcal{F}$ denotes the forward and $\mathcal{F}^{-1}$ denotes the inverse Fourier transform, $M$ is the operator that performs binary masking in k-space to with a given sampling mask. In Eq. (6) $\hat{k}_{Y^m_\Omega}$ and $\hat{Y}^m_\Omega$ denote undersampled multi-coil data respectively in k-space and image domain for the synthesized target contrast image. The selective loss function in ssGAN is then defined between undersampled synthesized and undersampled ground truth data for the target contrast, based on three loss components: multi-coil tensor image loss, multi-coil tensor k-space loss, and multi-coil tensor adversarial loss. Each loss term is described below.

\subsubsection{Multi-Coil Tensor Image Loss}
The first component of the selective loss function is a multi-coil tensor image loss defined based on undersampled multi-coil data in image domain, between synthesized and ground truth target images:
\eqal{}{
L_i=\mathrm{E}_{\textbf{X}_{\Lambda}^n,\textbf{Y}_{\Omega}^m}[||\hat{\textbf{Y}}^m_\Omega-\textbf{Y}^m_\Omega||_1]
}
where $\textbf{Y}_{\Omega}^m$ denotes the multi-coil ground truth target images from accelerated acquisitions, and $\hat{\textbf{Y}}^m_\Omega$ denotes the undersampled target images generated by ssGAN. 

\subsubsection{Multi-Coil Tensor k-space Loss}The quality of the synthesized images in ssGAN is further enhanced via a multi-coil tensor k-space loss expressed between the Fourier-domain data of the synthesized and ground truth images. 
\eqal{}{
L_k=\mathrm{E}_{\textbf{X}_{\Lambda}^n,\textbf{Y}_{\Omega}^m}[||h(\mathcal{F}(\hat{\textbf{Y}}^m_\Omega)/\beta)-h(\mathcal{F}(\textbf{Y}^m_\Omega)/\beta)||_1]
}
where $h$ is a $tanh$ function with a normalization constant $\beta$ to provide a comparable signal intensities across k-space, and $\mathcal{F}(\textbf{Y}^m_\Omega)$-$\mathcal{F}(\hat{\textbf{Y}}^m_\Omega)$ stand for k-space data of the ground truth and synthesized multi-coil images, respectively.

\subsubsection{Multi-Coil Tensor Adversarial Loss}
The level of realism in the synthesized images is advanced via a multi-coil adversarial loss function evaluated between image-domain data of the synthesized and ground truth multi-coil images:
\eqal{}{
L_{a}=-\mathrm{E}_{\textbf{Y}_{\Omega}^m}[(D(\textbf{Y}^m_\Omega)-1)^2]-\mathrm{E}_{\textbf{X}_{\Lambda}^n}[D(\hat{\textbf{Y}}^m_\Omega)^2]
}
where $D$ denotes the discriminator that distinguishes between undersampled ground truth and synthesized images. 

The final selective loss function for ssGAN is constructed as a weighted combination of the three multi-coil tensor loss terms described as $L_{ssGAN}=\lambda_kL_{k}+\lambda_iL_i+\lambda_{a}L_{a}$,
where $\lambda_k$, $\lambda_i$, and $\lambda_a$ denote the relative weighting of the tensor k-space, image, and adversarial losses. Note that the selective loss function in along with randomization of the k-space sampling masks across training subjects enables ssGAN to effectively capture complex relationships between acquired and non-acquired k-space coefficients. In turn, ssGAN can successfully recover high-quality target images without requiring Nyquist-sampled acquisitions of the target contrast. 

\subsection{Datasets}
The proposed ssGAN model was demonstrated on the public IXI dataset (https://brain-development.org/ixi-dataset/) containing multi-contrast single-coil magnitude brain images and an in-house dataset of multi-contrast multi-coil complex brain images.  
\subsubsection{The IXI Dataset}\Tone- and \Ttwo-weighted single-coil magnitude brain MR images of $94$ subjects were used, where $64$ were reserved for training, $10$ for validation, and $20$ for testing. See supplementary materials for scan parameters. Since multi-contrast images within subjects were spatially unaligned, \Ttwo-weighted images were registered onto \Tone-weighted images via FSL \cite{fsl} prior to experiments. Registration was performed using an affine transformation based on mutual information. For demonstrations, brain images for individual cross-sections were retrospectively undersampled in two-dimensions to yield acceleration ratios $R = [2:1:10]$, via uniform random sampling with a $10\times10$ central fully-sampled k-space region.

\subsubsection{In vivo Brain Dataset}\Ttwo-and PD-weighted multi-coil complex images of $10$ subjects were used, where $7$ were reserved for training, $1$ for validation, and $2$ for testing. See supplementary materials for scan parameters. Because there was negligible interscan motion, no spatial registration was performed. Data were collected on a $3T$ Siemens Magnetom scanner using a $32$-channel receive-only head coil at Bilkent University, Ankara, Turkey. Imaging protocols were approved by the local ethics committee at Bilkent University, and all participants provided written informed consent. To lower computational complexity, geometric-decomposition coil compression was performed to reduce the number of coils from $32$ to $5$ \cite{gcc}. For demonstrations, brain images for individual cross-sections were retrospectively undersampled in two-dimensions to yield acceleration ratios $R = [2:1:4]$, via uniform random sampling with a $16\times16$ central fully-sampled k-space region.

\subsection{Implementation Details}
The architecture of the generator and discriminator in ssGAN were adopted from a previous state-of-the-art study that demonstrated success in multi-contrast MRI synthesis \cite{dar2019}. The generator contained an encoder of $3$ convolutional layers, a residual network of $9$ ResNet blocks, and a decoder of $3$ convolutional layers in series. The discriminator contained a convolutional neural network of $5$ convolutional layers in series. An unlearned coil-combination block was placed at the input of the generator, so the generator recovered real-imaginary parts of the target image given real-imaginary parts of the coil-combined source image. The coil-combined target image was backprojected onto individual coils, and the complex target images from each coil were sequentially fed to an unconditional patch discriminator. The generator and the discriminator were alternately trained for $100$ epochs with a batch size of $1$ using the ADAM optimizer with first and second gradient moments of $\beta_1=0.5$ and $\beta_2=0.999$. The learning rate of the optimizer was set to $0.0002$ in the first $50$ epochs and was linearly decayed to $0$ in the last $50$ epochs. Cross-validation was used to select the relative weighting of the selective loss function components ($\lambda_i$, $\lambda_k$, $\lambda_a$, $\beta$) by maximizing synthesis performance based on network loss in the validation set. The set of parameters ($\lambda_i=100$, $\lambda_k=3000$, $\lambda_a=1$, $\beta=5000$) that yielded near-optimal performance in both datasets were used in all experiments. Implementations were run on nVidia 1080 Ti and 2080 Ti GPUs in Python2.7 using PyTorch. Code will be available at https://github.com/icon-lab/mrirecon.

\subsection{Competing Methods}
The proposed semi-supervised ssGAN model was comparatively demonstrated for multi-contrast MRI synthesis against several state-of-the-art methods. 

\subsubsection{pix2pix \cite{isola2017} (fully-sampled source, fully-sampled target)} The fully-supervised pix2pix model based on paired, Nyquist-sampled source-target acquisitions sets a gold-standard for synthesis performance. pix2pix learns a mapping between coil-combined, magnitude source-target images, so it is geared for single-coil MRI synthesis. Here, pix2pix was trained with single-coil nonselective variants of tensor losses in ssGAN. The generator-discriminator architectures were taken from \cite{dar2019}, and hyperparameters were optimized via cross-validation.

\subsubsection{CycleGAN \cite{zhu2017} (fully-sampled source, fully-sampled target)} The CycleGAN model based on unpaired albeit fully-sampled source-target acquisitions is another gold-standard reference. CycleGAN also learns the contrast mapping between coil-combined magnitude MR images, making it suitable for single-coil synthesis. The network architecture, hyperparameters, and loss functions (single-coil nonselective adversarial and cycle-consistency) in \cite{dar2019} were adopted.

\subsubsection{fsGAN (undersampled source, fully-sampled target)} The fsGAN model was constructed as a gold-standard supervised baseline in cases where target acquisitions were Nyquist sampled but source acquisitions were undersampled. fsGAN learns to map Fourier reconstructions of undersampled, multi-coil source acquisitions onto target images. Here it was trained using nonselective variants of tensor image, k-space and adversarial loss functions in ssGAN. The network architecture was matched to ssGAN. Hyperparameters were selected via cross-validation and identical to ssGAN.

\subsubsection{CasGAN (undersampled source, undersampled target)} CasGAN is a cascaded method that sequentially performs reconstruction and synthesis to cope with mutually accelerated source-target acquisitions. Here compressed-sensing reconstructions were first performed to recover source-target images from undersampled acquisitions. A fully-supervised synthesis model was then learned based on these reconstructions. CS reconstructions were implemented using SparseMRI \cite{SparseMRI} for single-coil data, and L$_1$-SPIRiT \cite{spirit} for multi-coil data (https://people.eecs.berkeley.edu/mlustig/Software.html). Hyperparameters were selected via cross-validation. In SparseMRI, the number of iterations was 4, weight for total variation regularization was $0.0001$ and weight for wavelet-domain L$_1$ regularization was $0.0001$. In SPIRiT, the kernel size was $5\times5$, weight for wavelet-domain L$_1$-regularization was 0.1, weight for Tikhonov regularization during kernel estimation was $0.001$, number of iterations was $10$ for PD-weighted images and 20 for \Ttwo-weighted images. The synthesis model had identical architecture, loss functions and hyperparameters to fsGAN.

\begin{figure}[htp]
\centerline{\includegraphics[width=0.40\textwidth]{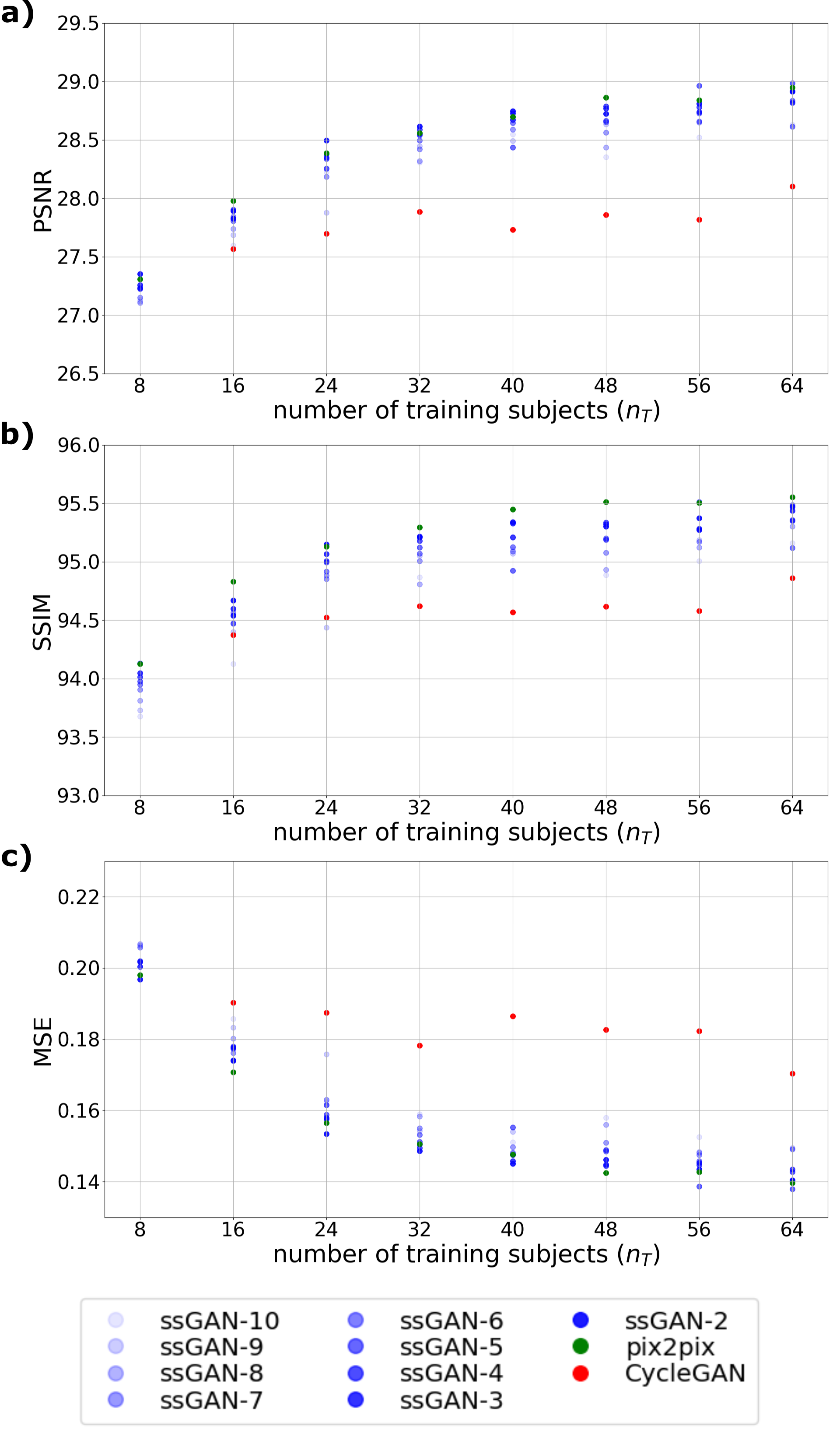}}
\caption{Reliability of ssGAN against training data deficiencies. Evaluations were performed for $n_T=[8:8:64]$. For each $n_T$, pix2pix and CycleGAN were trained with $R_{target}=1$, whereas ssGAN was trained with $R_{target}\in[2:1:10]$, ssGAN-$k$ with $k=R_{target}$. All models were trained with $R_{source}=1$. Performance metrics for CycleGAN at $n_T=8$ remain outside the display windows (see Fig. S1 for a broader display that also shows this model). }
\label{fig:robustness}
\end{figure}

\subsection{Experiments}
\subsubsection{Reliability against deficiencies in training data} Robustness against deficiencies in the quality and amount of training data was examined on the IXI dataset. Multiple independent ssGAN models were trained for \ToneTtwo~synthesis while varying the training dataset. Variations were introduced by altering the acceleration ratio of target contrast acquisitions across $R_{target}=[2:1:10]$, and by altering the number of training subjects across $n_T=[8:8:64]$. As gold-standard baselines, independent pix2pix and CycleGAN models were trained for the same variations in $n_T$ (albeit with $R_{target}=1$). Additional experiments were conducted on \TtwoTone~synthesis, where ssGAN models with $R_{target}=\{2,3,4\}$ were compared against pix2pix and CycleGAN. Fully-sampled source acquisitions $R_{source}=1$ were assumed for all experiments. 
\subsubsection{Single-coil synthesis} Experiments were conducted on brain images from the IXI dataset to demonstrate synthesis performance on single-coil data. Demonstrations were performed on the \ToneTtwo~and \TtwoTone~ synthesis tasks with ssGAN, fsGAN and CasGAN methods. Independent ssGAN and CasGAN models were trained for different target accelerations: ssGAN-$k$ and CasGAN-$k$ trained with $R_{target}=k$, where $k\in\{2,3,4\}$. ssGAN and all competing methods were separately trained for $R_{source}=\{2,3,4\}$. 

\subsubsection{Multi-coil synthesis} Experiments were conducted on brain images from the in vivo dataset to demonstrate synthesis performance on multi-coil data. Demonstrations were performed on the \TtwoPD~and \PDTtwo~ synthesis tasks with ssGAN, fsGAN and CasGAN. Independent ssGAN and CasGAN models were trained for different target accelerations: ssGAN-$k$ and CasGAN-$k$ trained with $R_{target}=k$, where $k\in\{2,3,4\}$. All competing methods were separately trained for $R_{source}=\{2,3,4\}$. A radiological evaluation was conducted on \TtwoPD~and \PDTtwo~synthesis tasks with $R_{source}=2,4$. Opinion scores of an expert radiologist with more than $25$ years of experience were considered. The quality of the synthesized images was rated based on similarity to reference images from fully-sampled acquisitions, on a five-point scale (0: unacceptable, 1: poor, 2: limited, 3: moderate, 4: good, 5: perfect match). For each synthesis task, radiological evaluations were performed on $5$ different cross-sections randomly taken from each subject. 

\subsubsection{Ablation studies} Experiments were conducted to individually examine the effects of the tensor image, k-space and adversarial loss functions on synthesis quality. Demonstrations were performed on IXI for \ToneTtwo~and \TtwoTone~synthesis tasks. Four independent ssGAN models were trained: ssGAN with all loss functions, ssGAN(w/o image) without the image loss, ssGAN(w/o k-space) without the k-space loss, and ssGAN(w/o adv) without the adversarial loss. Acceleration rates of $R_{source}=4$ and $R_{target}=4$ were assumed.

Synthesis performance was evaluated using peak signal-to-noise ratio (PSNR), structural similarity (SSIM) and mean-squared error (MSE) metrics (all MSE reports reflect measured MSE $\times$ 100). Metrics were measured on coil-combined magnitude images derived from synthesized and reference target contrasts. The reference image was based on Fourier reconstructions of fully-sampled target acquisitions. In Tables, summary statistics of quantitative metrics were provided as mean $\pm$ std across test subjects. Significance of PSNR, SSIM, MSE, and radiological opinions scores was assessed via Kruskal Wallis H-test ($p<0.05$) to collectively compare ssGAN models vs pix2pix, ssGAN models vs CycleGAN, and ssGAN models vs fsGAN, and via Wilcoxon signed-rank test ($p<0.05$) to individually compare ssGAN-2 vs CasGAN-2, ssGAN-3 vs CasGAN-3, and ssGAN-4 vs CasGAN-4.

\begin{figure}[t]
\centerline{\includegraphics[width=0.49\textwidth]{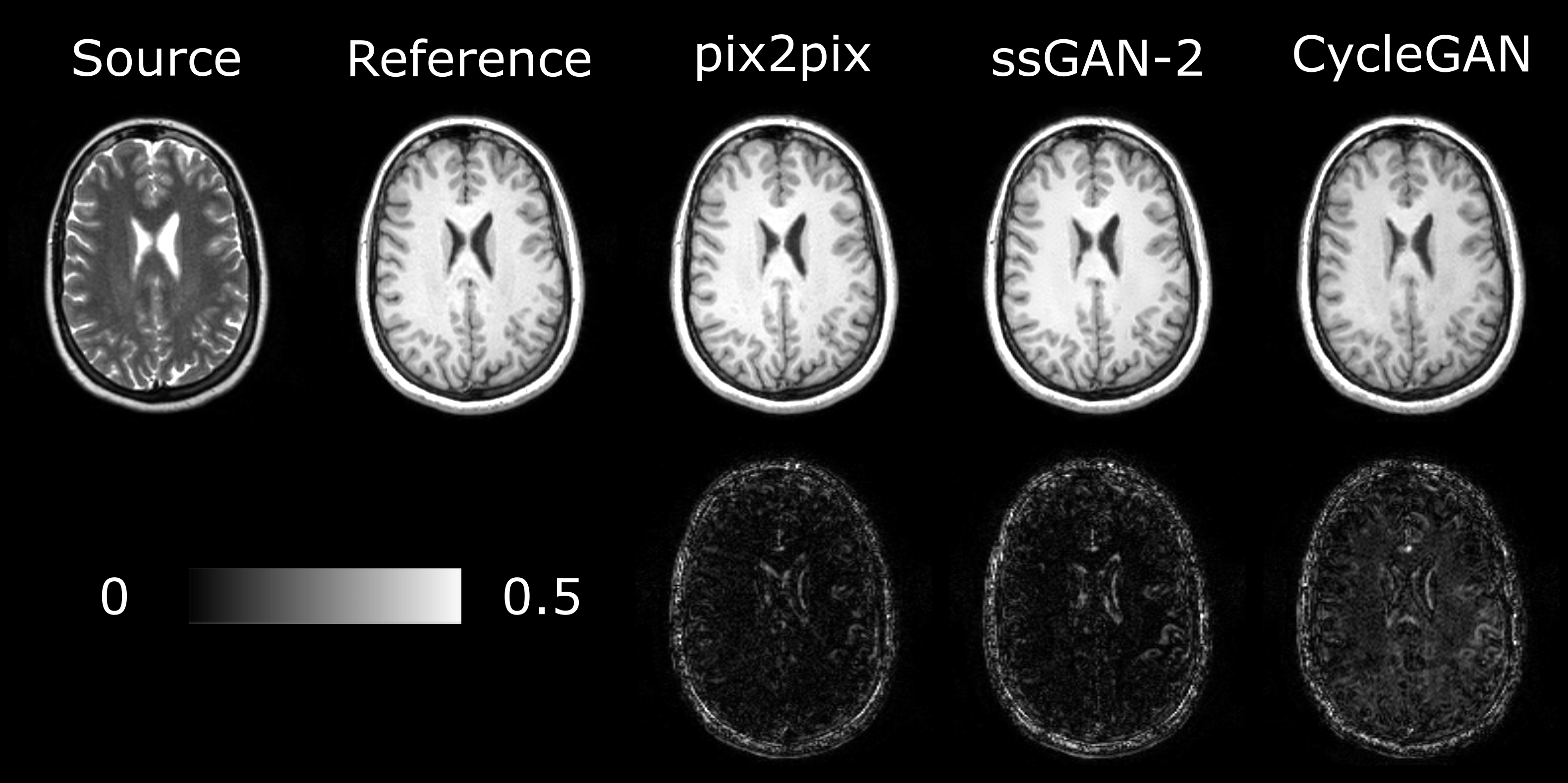}}
\caption{ssGAN was demonstrated on IXI for \TtwoTone~mapping against pix2pix and CycleGAN with ($R_{source}=1$). Synthesized images from ssGAN-2, pix2pix, and CycleGAN are displayed together with the reference (i.e., target) and source images in the first row. The corresponding error maps for the synthesized images are displayed in the second row (see colorbar). For comparison with ssGAN-3,-4, see Fig. S2.}
\label{fig:comp_cycpix}
\end{figure}

\begin{figure*}[htb]
\centerline{\includegraphics[width=0.97\textwidth]{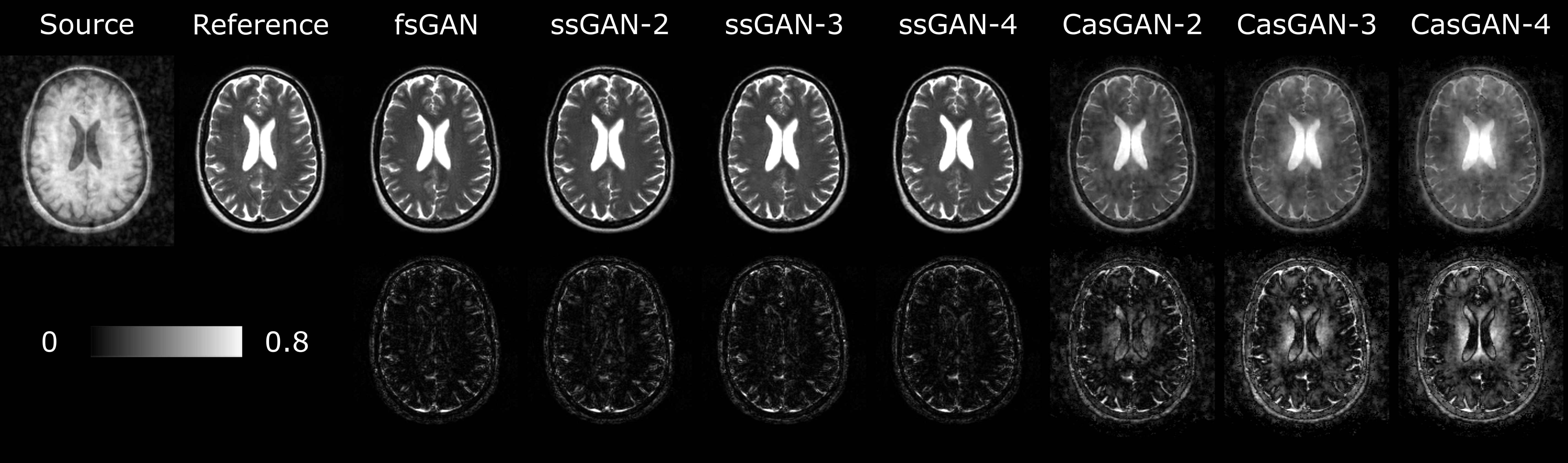}}
\caption{Synthesis quality of ssGAN, fsGAN and CasGAN was demonstrated on IXI for \ToneTtwo~synthesis ($R_{source}=2$). Synthesized images from the competing methods are displayed together with the source and reference (i.e. target) images in the first row, and the corresponding error maps for the synthesized images are displayed in the second row. }
\label{fig:single_coil}
\end{figure*}

\renewcommand{\tabcolsep}{3pt}
\renewcommand{\arraystretch}{1.30}
\begin{table*}[htb]
\caption{Image Quality for Mutually Acceleated Single-Coil MRI Synthesis}
\begin{subtable}[h]{0.25\textwidth}
\caption{$R_{source}=2$}
\scalebox{0.70}{
\begin{tabular}{ccccccc}
\cline{2-7}
& \multicolumn{3}{c}{\ToneTtwo} & \multicolumn{3}{c}{\TtwoTone} \\ \cline{2-7} 
& PSNR    	& SSIM     	& MSE    	& PSNR     	& SSIM     	& MSE    	\\ \hline
\multirow{2}{*}{fsGAN} &27.01 &93.63 &0.215 &27.54 &94.34 &0.194 \\
 &$\pm$1.41 &$\pm$1.53 &$\pm$0.072 &$\pm$1.45 &$\pm$1.51 &$\pm$0.067 \\ \hline
\multirow{2}{*}{ssGAN-2} &26.9 &93.36 &0.219 &27.46 &94.09 &0.196 \\
 &$\pm$1.37 &$\pm$1.52 &$\pm$0.072 &$\pm$1.35 &$\pm$1.53 &$\pm$0.061 \\ \hline
\multirow{2}{*}{ssGAN-3} &26.82 &93.21 &0.223 &27.3 &93.93 &0.206 \\
 &$\pm$1.38 &$\pm$1.53 &$\pm$0.074 &$\pm$1.47 &$\pm$1.6 &$\pm$0.072 \\ \hline
\multirow{2}{*}{ssGAN-4} &26.78 &93.13 &0.224 &27.29 &93.86 &0.204 \\
 &$\pm$1.34 &$\pm$1.5 &$\pm$0.072 &$\pm$1.38 &$\pm$1.58 &$\pm$0.066 \\ \hline
\multirow{2}{*}{CasGAN-2} &24.04 &85.22 &0.409 &21.25 &85.56 &0.84 \\
 &$\pm$0.81 &$\pm$2.12 &$\pm$0.079 &$\pm$1.05 &$\pm$2.31 &$\pm$0.176 \\ \hline
\multirow{2}{*}{CasGAN-3} &21.62 &81.46 &0.714 &18.07 &81.45 &1.672 \\
 &$\pm$0.72 &$\pm$2.43 &$\pm$0.122 &$\pm$0.8 &$\pm$2.58 &$\pm$0.287 \\ \hline
\multirow{2}{*}{CasGAN-4} &20.06 &79.03 &1.026 &16.81 &79.29 &2.167 \\
 &$\pm$0.6 &$\pm$2.35 &$\pm$0.141 &$\pm$0.69 &$\pm$2.68 &$\pm$0.345 \\ \hline

\end{tabular}
}
\end{subtable}
\hfil
\begin{subtable}[h]{0.25\textwidth}
\caption{$R_{source}=3$}
\scalebox{0.70}{
\begin{tabular}{ccccccc}
\cline{2-7}
& \multicolumn{3}{c}{\ToneTtwo} & \multicolumn{3}{c}{\TtwoTone} \\ \cline{2-7} 
& PSNR    	& SSIM     	& MSE    	& PSNR     	& SSIM     	& MSE    	\\ \hline
\multirow{2}{*}{fsGAN} &26.18 &92.58 &0.259 &27.01 &93.54 &0.215 \\
 &$\pm$1.4 &$\pm$1.63 &$\pm$0.089 &$\pm$1.27 &$\pm$1.54 &$\pm$0.065  \\ \hline
\multirow{2}{*}{ssGAN-2} &26.14 &92.22 &0.261 &27.05 &93.38 &0.214 \\
 &$\pm$1.35 &$\pm$1.61 &$\pm$0.087 &$\pm$1.31 &$\pm$1.58 &$\pm$0.065  \\ \hline
\multirow{2}{*}{ssGAN-3} &25.99 &92.06 &0.269 &26.91 &93.14 &0.22 \\
 &$\pm$1.32 &$\pm$1.57 &$\pm$0.087 &$\pm$1.3 &$\pm$1.59 &$\pm$0.066  \\ \hline
\multirow{2}{*}{ssGAN-4} &25.84 &91.69 &0.279 &26.83 &93.06 &0.225 \\
 &$\pm$1.34 &$\pm$1.65 &$\pm$0.093 &$\pm$1.3 &$\pm$1.61 &$\pm$0.067  \\ \hline
\multirow{2}{*}{CasGAN-2} &23.71 &84.35 &0.442 &21.16 &84.95 &0.853 \\
 &$\pm$0.84 &$\pm$2.23 &$\pm$0.089 &$\pm$1.05 &$\pm$2.3 &$\pm$0.19  \\ \hline
\multirow{2}{*}{CasGAN-3} &21.47 &80.82 &0.743 &17.97 &81.03 &1.705 \\
 &$\pm$0.74 &$\pm$2.49 &$\pm$0.131 &$\pm$0.86 &$\pm$2.62 &$\pm$0.312  \\ \hline
\multirow{2}{*}{CasGAN-4} &19.73 &78.08 &1.112 &16.91 &78.94 &2.126 \\
 &$\pm$0.67 &$\pm$2.48 &$\pm$0.172 &$\pm$0.68 &$\pm$2.72 &$\pm$0.338  \\ \hline

\end{tabular}
}
\end{subtable}
\hfil
\begin{subtable}[h]{0.25\textwidth}
\caption{$R_{source}=4$}
\scalebox{0.70}{
\begin{tabular}{ccccccc}
\cline{2-7}
                           & \multicolumn{3}{c}{\ToneTtwo} & \multicolumn{3}{c}{\TtwoTone} \\ \cline{2-7} 
                           			& PSNR    	& SSIM     	& MSE    	& PSNR     	& SSIM     	& MSE    	\\ \hline
\multirow{2}{*}{fsGAN} &25.51 &91.61 &0.302 &26.81 &93.1 &0.227 \\
 &$\pm$1.39 &$\pm$1.71 &$\pm$0.103 &$\pm$1.29 &$\pm$1.64 &$\pm$0.068 \\ \hline
\multirow{2}{*}{ssGAN-2} &25.46 &91.27 &0.305 &26.63 &92.75 &0.236 \\
 &$\pm$1.38 &$\pm$1.73 &$\pm$0.104 &$\pm$1.29 &$\pm$1.64 &$\pm$0.069 \\ \hline
\multirow{2}{*}{ssGAN-3} &25.35 &91.13 &0.313 &26.54 &92.59 &0.241 \\
 &$\pm$1.37 &$\pm$1.75 &$\pm$0.109 &$\pm$1.32 &$\pm$1.68 &$\pm$0.072 \\ \hline
\multirow{2}{*}{ssGAN-4} &25.26 &90.79 &0.321 &26.5 &92.32 &0.243 \\
 &$\pm$1.43 &$\pm$1.83 &$\pm$0.114 &$\pm$1.28 &$\pm$1.73 &$\pm$0.072 \\ \hline
\multirow{2}{*}{CasGAN-2} &23.49 &83.7 &0.465 &21.04 &84.42 &0.88 \\
 &$\pm$0.9 &$\pm$2.31 &$\pm$0.104 &$\pm$1.19 &$\pm$2.39 &$\pm$0.213 \\ \hline
\multirow{2}{*}{CasGAN-3} &21.3 &80.07 &0.773 &18.01 &80.66 &1.707 \\
 &$\pm$0.75 &$\pm$2.58 &$\pm$0.134 &$\pm$0.89 &$\pm$2.69 &$\pm$0.329 \\ \hline
\multirow{2}{*}{CasGAN-4} &19.85 &77.72 &1.078 &16.87 &78.55 &2.142 \\
 &$\pm$0.66 &$\pm$2.52 &$\pm$0.169 &$\pm$0.68 &$\pm$2.73 &$\pm$0.347 \\ \hline

\end{tabular}
}
\end{subtable}

\end{table*}

\section{Results}

\subsection{Robustness of semi-supervised learning against deficiencies in training data}

We first performed comprehensive experiments to examine the reliability of ssGAN against deficiencies in training data. Models were learned for \ToneTtwo~mapping in the IXI dataset, while acceleration ratio of target acquisitions ranged in $R_{target}=[2:1:10]$ and number of training subjects ranged in $n_T=[8:8:64]$. As baselines, gold-standard pix2pix and CycleGAN models were also trained for matching $n_T$ but with $R_{target}=1$. All models were input high-quality source images reconstructed from Nyquist-sampled acquisitions ($R_{source}=1$). Synthesis performance in terms of PSNR, SSIM, and MSE is displayed in Fig. \ref{fig:robustness} as a function of $R_{target}$ and $n_T$. The reported measurements indicate that synthesis quality of ssGAN is on par with the gold-standard pix2pix model ($p>0.05$), where ssGAN performance is within $[-0.51,0.13]$ dB PSNR, $[-0.70,0.02]$ \% SSIM, and $[0.019,-0.004]$ MSE windows of pix2pix. Meanwhile, ssGAN outperforms CycleGAN with $1.18$ dB higher PSNR, $1.16$ \% SSIM, and $0.081$ lower MSE ($p<0.05$). Importantly, ssGAN models trained with varying $R_{target}$ yield highly similar performance, where ssGAN achieves near-optimal synthesis quality while undersampling target acquisitions up to $10$-fold.

Measurements reported in Fig. \ref{fig:robustness} also demonstrate that increasing $n_T$ improves performance of all competing methods. Comparing $n_T=16$ against $n_T=32$, average improvements in (PSNR, SSIM, MSE) are ($0.68\,\mathrm{dB},0.57\,\%,-0.025$) for ssGAN, ($0.58\,\mathrm{dB}, 0.47\,\%, -0.020$) for pix2pix, and ($0.32\,\mathrm{dB},0.24\,\%,-0.012$) for CycleGAN. Note that ssGAN improves scan efficiency by accelerating target acquisitions, so in principle training data from a larger group of subjects can be collected at high acceleration rates for training ssGAN, compared to fully-supervised or unpaired models. For instance, given a total, active scan time of $126$ min, Nyquist-sampled k-space data for \Tone- and \Ttwo-weighted images can be collected in $16$ subjects for pix2pix and CycleGAN. In the same duration, a protocol with undersampled target acquisitions ($R_{target}=10$) can be performed in 32 subjects for ssGAN, resulting in performance benefits of ($0.34\,\mathrm{dB},0.04\,\%,-0.012$) over pix2pix and ($0.74\,\mathrm{dB},0.48\,\%,-0.031$) over CycleGAN. Therefore, ssGAN enables elevated diversity in the training set to improve accuracy and practicality of learning-based MRI synthesis.

\begin{figure*}[]
\centerline{\includegraphics[width=0.99\textwidth]{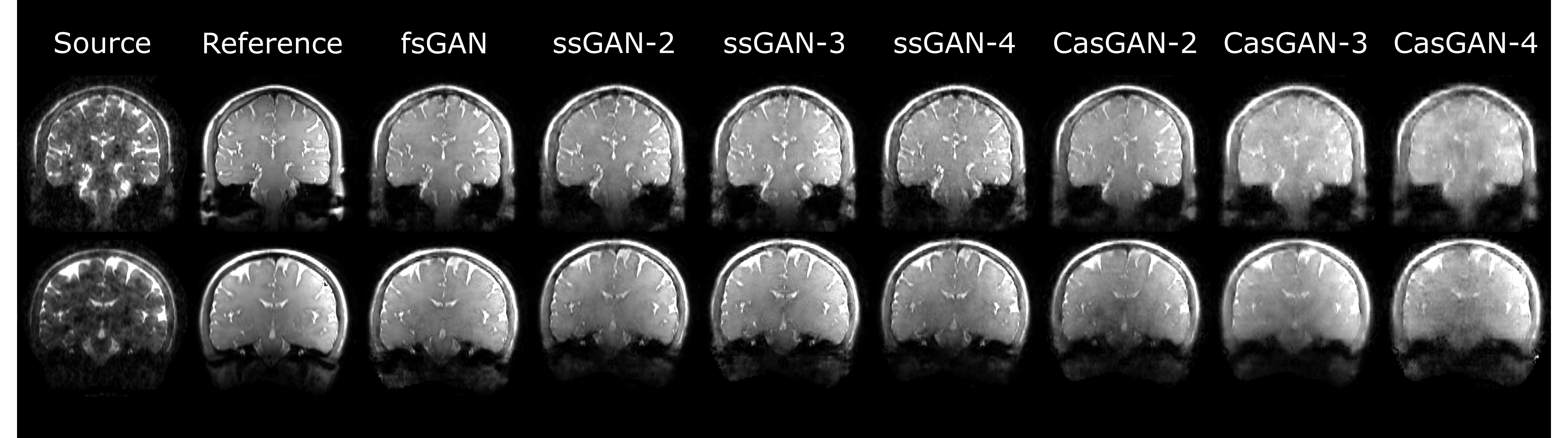}}
\caption{Synthesis quality of ssGAN, fsGAN and CasGAN was demonstrated on the in vivo dataset for \TtwoPD~synthesis ($R_{source}=2$). Representative results from two different subjects are displayed, along with the source and reference images.}
\label{fig:multi_coil}
\end{figure*}

\renewcommand{\tabcolsep}{3pt}
\renewcommand{\arraystretch}{1.30}
\begin{table*}[ht]
\caption{Image Quality for Mutually Accelerated Multi-Coil MRI Synthesis}
\begin{subtable}[h]{0.25\textwidth}
\caption{$R_{source}=2$}
\scalebox{0.7}{
\begin{tabular}{ccccccc}
\cline{2-7}
                           & \multicolumn{3}{c}{\TtwoPD} & \multicolumn{3}{c}{\PDTtwo} \\ \cline{2-7} 
                           			& PSNR    	& SSIM     	& MSE    	& PSNR     	& SSIM     	& MSE    	\\ \hline

\multirow{2}{*}{fsGAN} &25.48 &87.8 &0.295 &25.35 &87.78 &0.306 \\
 &$\pm$0.13 &$\pm$1.46 &$\pm$0.015 &$\pm$0.46 &$\pm$1.63 &$\pm$0.032 \\ \hline
\multirow{2}{*}{ssGAN-2} &25.75 &88.14 &0.279 &24.59 &86.56 &0.361 \\
 &$\pm$0.01 &$\pm$1.83 &$\pm$0.005 &$\pm$0.21 &$\pm$0.52 &$\pm$0.016 \\ \hline
\multirow{2}{*}{ssGAN-3} &25.16 &87.42 &0.315 &24.59 &86.25 &0.359 \\
 &$\pm$0.26 &$\pm$0.97 &$\pm$0.022 &$\pm$0.13 &$\pm$0.42 &$\pm$0.01 \\ \hline
\multirow{2}{*}{ssGAN-4} &25.47 &87.72 &0.296 &24.46 &85.97 &0.37  \\
 &$\pm$0.15 &$\pm$1.92 &$\pm$0.006 &$\pm$0.2 &$\pm$0.3 &$\pm$0.017 \\ \hline
\multirow{2}{*}{CasGAN-2} &25.01 &86.96 &0.347 &24.44 &85.02 &0.466 \\
 &$\pm$0.03 &$\pm$1.61 &$\pm$0.021 &$\pm$0.43 &$\pm$2.45 &$\pm$0.102 \\ \hline
\multirow{2}{*}{CasGAN-3} &23.91 &84.72 &0.443 &24.23 &83.54 &0.465 \\
 &$\pm$0.15 &$\pm$1.76 &$\pm$0.042 &$\pm$0.81 &$\pm$3.74 &$\pm$0.137 \\ \hline
\multirow{2}{*}{CasGAN-4} &22.95 &82.22 &0.562 &23.55 &81.53 &0.487 \\
 &$\pm$0.59 &$\pm$2.32 &$\pm$0.112 &$\pm$0.24 &$\pm$3.03 &$\pm$0.061 \\ \hline

\end{tabular}
}
\end{subtable}
\hfil
\begin{subtable}[h]{0.25\textwidth}
\caption{$R_{source}=3$}
\scalebox{0.7}{
\begin{tabular}{ccccccc}
\cline{2-7}
                          & \multicolumn{3}{c}{\TtwoPD} & \multicolumn{3}{c}{\PDTtwo} \\ \cline{2-7} 
                           			& PSNR    	& SSIM     	& MSE    	& PSNR     	& SSIM     	& MSE    	\\ \hline
\multirow{2}{*}{fsGAN} &25.04 &86.5 &0.338 &24.77 &86.73 &0.345 \\
 &$\pm$0.42 &$\pm$0.56 &$\pm$0.048 &$\pm$0.07 &$\pm$0.79 &$\pm$0.003 \\ \hline
\multirow{2}{*}{ssGAN-2} &25.14 &87.08 &0.315 &24.65 &86.37 &0.356 \\
 &$\pm$0.01 &$\pm$1.61 &$\pm$0.003 &$\pm$0.02 &$\pm$0.56 &$\pm$0.006 \\ \hline
\multirow{2}{*}{ssGAN-3} &25.2 &87.15 &0.311 &24.61 &86.09 &0.358 \\
 &$\pm$0.05 &$\pm$1.37 &$\pm$0.005 &$\pm$0.06 &$\pm$0.67 &$\pm$0.003 \\ \hline 
\multirow{2}{*}{ssGAN-4} &25.01 &86.78 &0.325 &24.16 &85.25 &0.398 \\
 &$\pm$0.19 &$\pm$1.25 &$\pm$0.018 &$\pm$0.12 &$\pm$0.19 &$\pm$0.016 \\ \hline
\multirow{2}{*}{CasGAN-2} &24.77 &86.41 &0.374 &24.57 &82.86 &0.428 \\
 &$\pm$0.32 &$\pm$1.81 &$\pm$0.051 &$\pm$1.44 &$\pm$6.67 &$\pm$0.169 \\ \hline
\multirow{2}{*}{CasGAN-3} &23.44 &83.9 &0.496 &24.2 &82.11 &0.444 \\
 &$\pm$0.36 &$\pm$2.05 &$\pm$0.073 &$\pm$1.24 &$\pm$6.18 &$\pm$0.153 \\ \hline
\multirow{2}{*}{CasGAN-4} &22.68 &81.76 &0.579 &22.83 &77.46 &0.59 \\
 &$\pm$0.04 &$\pm$1.8 &$\pm$0.037 &$\pm$1.0 &$\pm$5.75 &$\pm$0.164 \\ \hline

\end{tabular}
}
\end{subtable}
\hfil
\begin{subtable}[h]{0.25\textwidth}
\caption{$R_{source}=4$}
\scalebox{0.7}{
\begin{tabular}{ccccccc}
\cline{2-7}
                           & \multicolumn{3}{c}{\TtwoPD} & \multicolumn{3}{c}{\PDTtwo} \\ \cline{2-7} 
                           			& PSNR    	& SSIM     	& MSE    	& PSNR     	& SSIM     	& MSE    	\\ \hline
\multirow{2}{*}{fsGAN} &24.9 &86.51 &0.337 &24.51 &86.16 &0.365 \\
 &$\pm$0.06 &$\pm$1.9 &$\pm$0.003 &$\pm$0.33 &$\pm$0.07 &$\pm$0.031 \\ \hline
\multirow{2}{*}{ssGAN-2} &24.85 &86.64 &0.338 &24.35 &85.66 &0.379 \\
 &$\pm$0.0 &$\pm$1.47 &$\pm$0.002 &$\pm$0.35 &$\pm$0.2 &$\pm$0.034 \\ \hline
\multirow{2}{*}{ssGAN-3} &24.6 &86.12 &0.359 &24.3 &85.5 &0.383 \\
 &$\pm$0.08 &$\pm$1.74 &$\pm$0.006 &$\pm$0.37 &$\pm$0.32 &$\pm$0.036 \\ \hline
\multirow{2}{*}{ssGAN-4} &24.73 &86.22 &0.35 &24.05 &84.84 &0.406 \\
 &$\pm$0.12 &$\pm$1.83 &$\pm$0.008 &$\pm$0.38 &$\pm$0.39 &$\pm$0.039 \\ \hline
\multirow{2}{*}{CasGAN-2} &23.72 &84.78 &0.479 &25.12 &85.91 &0.353 \\
 &$\pm$0.48 &$\pm$2.24 &$\pm$0.085 &$\pm$1.33 &$\pm$3.81 &$\pm$0.124 \\ \hline
\multirow{2}{*}{CasGAN-3} &23.09 &82.59 &0.553 &24.84 &84.83 &0.374 \\
 &$\pm$0.41 &$\pm$2.37 &$\pm$0.099 &$\pm$1.29 &$\pm$3.87 &$\pm$0.127 \\ \hline
\multirow{2}{*}{CasGAN-4} &22.31 &80.32 &0.663 &23.51 &81.68 &0.491 \\
 &$\pm$0.23 &$\pm$2.5 &$\pm$0.095 &$\pm$1.17 &$\pm$3.79 &$\pm$0.145 \\ \hline

\end{tabular}
}
\end{subtable}

\end{table*}

We then extended the demonstrations of ssGAN against pix2pix and CycleGAN by comparisons on the \TtwoTone~synthesis task in the IXI dataset with a fixed number of training subjects $n_T=32$ (used hereafter in all evaluations in IXI). Measurements of synthesis quality are reported in Table S1. The reported measurements reveal that ssGAN models maintain near-optimal synthesis quality on par with pix2pix, and on average they outperform CycleGAN with ($0.41\,\mathrm{dB}, 0.34\,\%,-0.015$) improvement in (PSNR, SSIM, MSE). Representative results displayed in Fig. \ref{fig:comp_cycpix} corroborate the quantitative findings by showing that ssGAN offers a similar level of accuracy in tissue depiction to pix2pix, while it synthesizes higher quality images compared to CycleGAN that suffers from elevated errors.

\subsection{Single-coil image synthesis in mutually accelerated multi-contrast MRI}
To examine the synthesis performance of ssGAN in mutually accelerated MRI, we conducted experiments on IXI where both source and target acquisitions were undersampled. Single-coil image synthesis was considered with \ToneTtwo~and \TtwoTone~recovery tasks. ssGAN was compared against a gold-standard supervised model (fsGAN) that was trained on undersampled source acquisitions but Nyquist-sampled target acquisitions, and against a sequential model (CasGAN) that first reconstructed undersampled acquisitions, and then trained a synthesis model on the reconstructed source and target images. The target acceleration ratio varied in $R_{target} = \{2,3,4\}$ for ssGAN and CasGAN resulting in three separate models for each method: ssGAN-$k$ and CasGAN-$k$ with $k=R_{target}$. Meanwhile, the acceleration ratio for the source acquisitions varied in $R_{source} = \{2,3,4\}$ for all methods.

Quantitative metrics for synthesis performance  are listed in Table I for varying $R_{source}$ and $R_{target}$ values. Overall, ssGAN models at moderate acceleration factors for the target acquisition yield near-optimal performance on par with the reference fsGAN model ($p>0.05$), while mitigating the demands for Nyquist-sampled target acquisitions. Furthermore, ssGAN outperforms CasGAN by an average of $6.32$ dB in PSNR, $11.26$ \% in SSIM, and $-0.914$ in MSE ($p<0.05$). On average, incremental steps from $R_{target}=1$ to $R_{target}=4$ result in modest performance losses of $0.10$ dB PSNR, $0.19$ \% SSIM and $0.005$ MSE for ssGAN. In contrast, CasGAN suffers from elevated losses of $2.04$ dB PSNR, $3.05$ \% SSIM and $0.805$ MSE. This finding demonstrates that the selective loss function in ssGAN effectively copes with the reduction in quality of target acquisitions. 

Representative synthesis results from the methods under comparison are shown in Fig. \ref{fig:single_coil}. Quality of synthetic images from ssGAN are virtually identical to those from the supervised fsGAN model, and they are superior to the CasGAN model that suffers from residual artifacts and noise that carry over from the initial reconstruction stage.

\subsection{Multi-coil image synthesis in mutually accelerated multi-contrast MRI}
Next, we conducted experiments on the in vivo brain dataset to demonstrate multi-coil MRI synthesis with the proposed ssGAN model. Multi-coil image synthesis was considered for \TtwoPD~and \PDTtwo~recovery tasks. As in single-coil synthesis, ssGAN was compared against fsGAN and CasGAN. $R_{target} = \{2,3,4\}$ and $R_{source} = \{2,3,4\}$ were considered. 

Quantitative measurements for synthesis quality are reported in Table II for various $R_{source}$ and $R_{target}$ values. Overall, ssGAN models at distinct acceleration factors for the target acquisitions yield near-optimal performance on par with the gold-standard fsGAN model ($p>0.05$). On average across $R_{target}$, ssGAN outperforms CasGAN by $0.92$ dB in PSNR, $3.18$ \% in SSIM, and $-0.130$ in MSE ($p<0.05$, except for \PDTtwo~with $R_{source}=4$). Incremental steps from $R_{target}=1$ to $R_{target}=4$ result in an average performance loss of $0.12$ dB PSNR, $0.30$ \% SSIM, and $0.010$ MSE for ssGAN, and $0.82$ dB PSNR, $2.25$ \% SSIM, and $0.07$ MSE for CasGAN. Similar to single-coil results, this finding demonstrates the utility of the selective loss function in ssGAN to cope with moderately undersampled target acquisitions.

For further validation of the quantitative assessments, radiological evaluations were performed for ssGAN-2, fsGAN and CasGAN-2 on \TtwoPD~and \PDTtwo~synthesis tasks. Representative synthetic images are displayed in Fig. 5 and Fig. S3, whereas results of radiological evaluation are shown in Fig. 6. ssGAN images are visually similar to fsGAN, whereas they manifest superior synthesis quality compared to CasGAN. In terms of opinion score, ssGAN maintains a high-level of synthesis quality on par with fsGAN ($p>0.05$, except for $R_{source}=2$), and on average a modest score difference of $0.325$ is observed. In contrast, ssGAN yields superior performance to CasGAN with an average improvement of $1.075$ in opinion score across tasks ($p<0.05$).

\subsection{Ablation Studies}
Ablation experiments were conducted to demonstrate the contribution of individual loss components in ssGAN. Independent ssGAN models were trained while the loss components were selectively ablated (see Experiments). Models were learned for \ToneTtwo~and \TtwoTone~mappings in the IXI dataset. The effects of image and k-space losses were evaluated using PSNR, SSIM and MSE metrics, whereas the effect of adversarial loss was assessed using Frechlet Inception Distance (FID) scores and visual inspection as common in literature \cite{fid}. Quantitative metrics listed in Table S2 indicate that the selective image and k-space losses serve to improve synthesis quality in both \Tone~and \Ttwo~recovery tasks. Meanwhile, the selective adversarial loss component increases the realism of synthetic images with decreased FID scores.

\section{Discussion}
Here we introduced a novel semi-supervised deep generative model for image synthesis in multi-contrast MRI that is mutually accelerated across both contrast sets and k-space. As opposed to supervised models \cite{chartsias2017,sevetlidis2016,joyce2017,wei2019,bowles2016,dar2019,sharma2019}, ssGAN learns to synthesize high-quality target-contrast images in the absence of training sets composed of costly acquisitions of Nyquist-sampled source and target contrasts. ssGAN achieves synthesis quality on par with gold-standard supervised models for a broad range of acceleration ratios. This performance leap is mediated by selective loss functions in image, k-space, and adversarial domains. Unlike prior synthesis methods, ssGAN processes multi-coil complex MRI data and learns to synthesize directly from undersampled source acquisitions. Therefore, ssGAN holds great promise in advancing the practicality and utility of multi-contrast MRI synthesis.

\begin{figure}[]
\centerline{\includegraphics[width=0.49\textwidth]{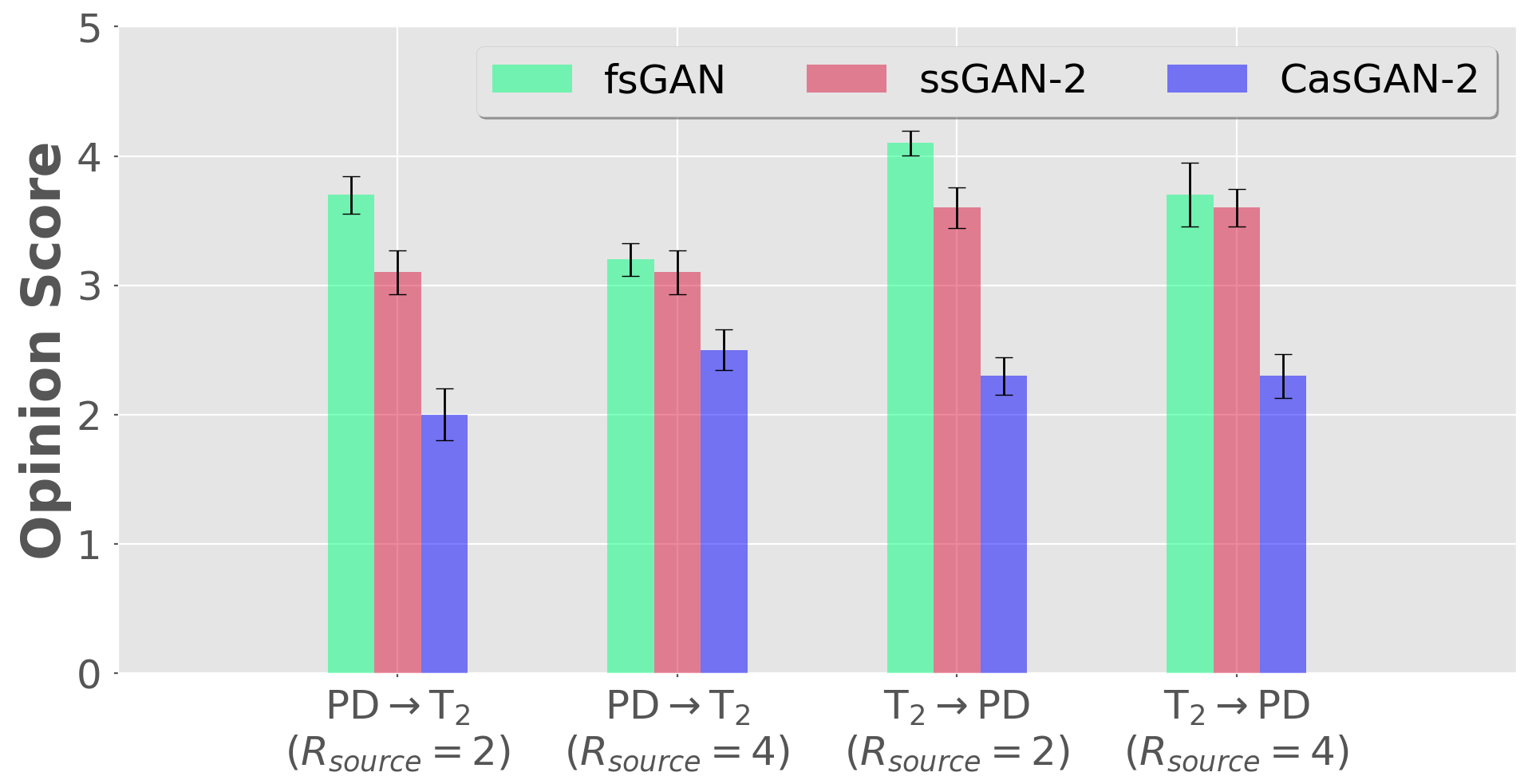}}
\caption{Radiological evaluations for fsGAN, ssGAN-2, and CasGAN-2 are shown. \TtwoPD~and \PDTtwo~synthesis tasks at $R_{source}=2,4$ were assessed on the in vivo dataset.}
\label{fig:multi_coil}
\end{figure}

Comprehensive experiments were conducted on single- and multi-coil neuroimaging datasets to demonstrate the proposed approach. Our experiments indicate that ssGAN achieves equivalent performance to gold-standard fully-supervised models, pix2pix \cite{zhu2017} with fully-sampled source acquisitions, and fsGAN with undersampled source acquisitions. Importantly, ssGAN maintains near-optimal synthesis for acceleration factors up to 10 for target acquisitions given the same amount of training subjects, significantly reducing the data requirements for model training. Furthermore, ssGAN outperforms an alternative weakly-supervised synthesis model CasGAN based on a sequential hybrid of compressed-sensing reconstructions followed by supervised synthesis. Compared to CasGAN, ssGAN enables end-to-end learning of target contrast recovery, alleviating potential propagation of errors across the pipeline and reducing computational complexity. 

Here, we demonstrated ssGAN with uniform-density random undersampling in k-space. An alternative would be to collect low-resolution source-target images by Nyquist-sampling within a central k-space region to achieve similar acceleration. The problem would then be transformed into a superresolution task on coil-combined images \cite{superres1,superres2}. However,  this superresolution task is challenging when both source and target images are low resolution, so external priors might be required to enable recovery of high-spatial-frequency information. Another alternative would be to perform variable-density sampling where central k-space is more densely sampled \cite{SparseMRI}. Variable-density schemes trade-off high-frequency coverage in return for increased signal-to-noise ratio in acquired data. As such, they might improve aggregate performance metrics (e.g., PSNR, MSE) that are dominated by low-spatial-frequency errors, particularly for higher acceleration factors used in ssGAN and CasGAN. Note, however, that uniform-density sampling improves peripheral k-space coverage to expand high-spatial-frequency information, and in turn contributes to recovery of detailed tissue structure. 

In the current study, we demonstrated the proposed method on one-to-one synthesis tasks with a single source and a single target contrast. In multi-contrast protocols, multiple source and/or target contrasts might be available, one might additionally be interested in many-to-one and many-to-many synthesis tasks \cite{sharma2019,lee2019colla}. In such cases, a many-to-many variant of ssGAN can be constructed by concatenating the multitude of source and target contrasts as separate input and output channels, respectively \cite{sharma2019}. The selective loss function along with the k-space masking and coil projection operators can then be defined separately for each target contrast.

The ssGAN implementation considered here leverages a multi-coil tensor loss between undersampled versions of the synthesized and reference target images. This image-domain loss implicitly assumes that the source and target-contrast acquisitions are spatially registered. The datasets examined here were either aligned or a registration step was performed during preprocessing. If an end-to-end alternative is desired that can cope with misaligned source-target acquisitions, deep network-based registration models can be cascaded to the input of ssGAN for spatial registration \cite{registration}. It remains important future work to explore the extent of improvements in synthesis performance with integrated registration and synthesis. 

The semi-supervised learning framework that ssGAN leverages undersampled albeit paired acquisitions of source and target contrasts from the same set of subjects. Our results suggest that successful ssGAN models can be trained even with relatively modest size datasets. However, more complex models including order of magnitude higher number of parameters such as 3D architectures might require substantial datasets for reliable training. In such cases, a variant of ssGAN that permits training on a hybrid of paired and unpaired images or directly on unpaired images would be valuable. To do this, the cycle-consistent counterpart of the selective loss function in ssGAN can be devised \cite{dar2019,lee2019colla,olut2018,jin2018,wang2020semi}.    

In summary, here we proposed a semi-supervised learning framework based on generative adversarial networks that can recover high-quality target images without demanding Nyquist-sampled ground truths. While the superior data-efficiency of ssGAN was primarily demonstrated for within-modality contrast conversion in the brain, it can also be adopted to other anatomies, other recovery tasks including multi-parametric MRI synthesis, or cross-modality mappings between MRI and other imaging modalities \cite{jin2018,wolterink2017,dewey2019}.

\bibliographystyle{ieeetr}
\bibliography{ssganbib}

\onecolumn

\clearpage
\newpage
\setcounter{page}{1}

\captionsetup[table]{name= Supp. Table}
\setcounter{table}{0} \renewcommand{\thetable}{\arabic{table}}

\setcounter{figure}{0} \renewcommand{\thefigure}{Supp. \arabic{figure}}

\setcounter{table}{0}
\renewcommand{\thetable}{S\arabic{table}}   
\setcounter{figure}{0}
\renewcommand{\thefigure}{S\arabic{figure}}%

\subsection*{Supp. Text}
\noindent Acquisition Parameters of the IXI Dataset
\begin{itemize}
\item \textbf{\Tone-weighted images:} TR $=9.81\,ms$, TE $=4.603\,ms$, flip angle $=8^\circ$, matrix size $=256\times256\times150$, spatial resolution $=0.94\times0.94\times1.2\,mm^3$, acquisition time $=4:42$.
\item \textbf{\Ttwo-weighted images:} TR $=8178.34\,ms$, TE $=100\,ms$, flip angle $=90^\circ$, matrix size $=256\times256\times150$, spatial resolution $=0.94\times0.94\times1.2\,mm^3$, acquisition time $=3:11$.
\end{itemize}

\noindent Acquisition Parameters of the In Vivo Brain Dataset
\begin{itemize}
\item \textbf{\Ttwo-weighted images:} 3D spin-echo sequence, TR $=1000\,ms$, TE $=118\,ms$, flip angle $=90^\circ$, imaging matrix $=256\times192\times88$, spatial resolution $=1\times1\times2\,mm^3$, acquisition time $=17:39$.
\item \textbf{PD-weighted images:} 3D spin-echo sequence, TR $=750\,ms$, TE $=12\,ms$, flip angle $=90^\circ$, imaging matrix $=256\times192\times88$, spatial resolution $=1\times1\times2\,mm^3$, acquisition time $=13:14$. 
\end{itemize}

\clearpage
\newpage
\begin{figure}[htp]
\centerline{\includegraphics[width=0.55\textwidth]{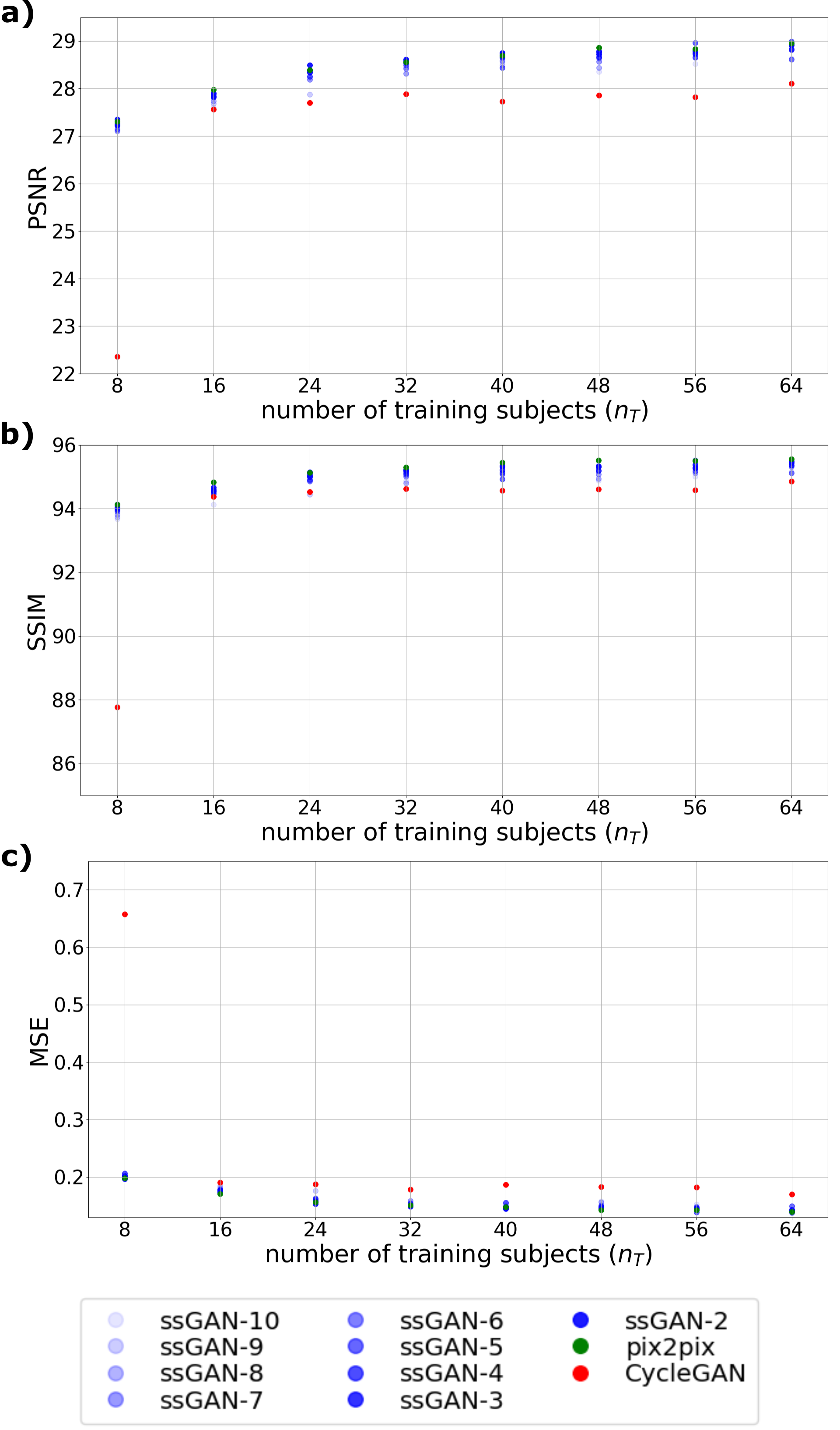}}
\caption{Reliability of ssGAN against training data deficiencies. Evaluations were performed for $n_T=[8:8:64]$. For each $n_T$, pix2pix and CycleGAN were trained with $R_{target}=1$, whereas ssGAN was trained with $R_{target}\in[2:1:10]$, ssGAN-$k$ with $k=R_{target}$. All models were trained with $R_{source}=1$.}
\label{fig:robustness_supp}
\end{figure}

\clearpage
\newpage
\renewcommand{\tabcolsep}{3pt}
\renewcommand{\arraystretch}{1.40}
\begin{table*}[h]
\begin{center}
\caption{Quality of Synthesis in the IXI Dataset for Nyquist-Sampled Source Acquisitions}
\scalebox{0.99}{
\begin{tabular}{ccccccc}
\cline{2-7}
                           & \multicolumn{3}{c}{\ToneTtwo} & \multicolumn{3}{c}{\TtwoTone} \\ \cline{2-7} 
                           			& PSNR    		& SSIM     		& MSE    		& PSNR     	& SSIM     		& MSE    	\\ \hline
\multirow{2}{*}{pix2pix} 		& 28.57 		& 95.33		& 0.15		& 28.62		& 95.95		& 0.154		\\
 						& $\pm$ 1.39	& $\pm$ 1.35	& $\pm$ 0.048	& $\pm$ 1.45	& $\pm$ 1.38	& $\pm$ 0.057	\\ \hline
\multirow{2}{*}{CycleGAN} 		& 27.91		& 94.66		& 0.178		& 28.05		& 95.4		& 0.175		\\
 						& $\pm$ 1.52	& $\pm$ 1.51	& $\pm$ 0.062	& $\pm$ 1.41	& $\pm$ 1.52	& $\pm$ 0.062	\\ \hline
\multirow{2}{*}{ssGAN-2} 		& 28.63		& 95.25		& 0.148		& 28.52		& 95.79		& 0.158		\\
 						& $\pm$ 1.42	& $\pm$ 1.33	& $\pm$ 0.049	& $\pm$ 1.61	& $\pm$ 1.46	& $\pm$ 0.061	\\ \hline
\multirow{2}{*}{ssGAN-3} 		& 28.56		& 95.24		& 0.15		& 28.44		& 95.72		& 0.16		\\
 						& $\pm$ 1.39	& $\pm$ 1.32	& $\pm$ 0.048	& $\pm$ 1.55	& $\pm$ 1.41	& $\pm$ 0.059	\\ \hline
\multirow{2}{*}{ssGAN-4} 		& 28.6		& 95.21		& 0.149		& 28.42		& 95.7		& 0.162		\\
 						& $\pm$ 1.43	& $\pm$ 1.34	& $\pm$ 0.05	& $\pm$ 1.56	& $\pm$ 1.42	& $\pm$ 0.061	\\ \hline

\end{tabular}
}
\end{center}
\end{table*}

\clearpage
\newpage
\begin{figure*}[htp]
\centerline{\includegraphics[width=0.89\textwidth]{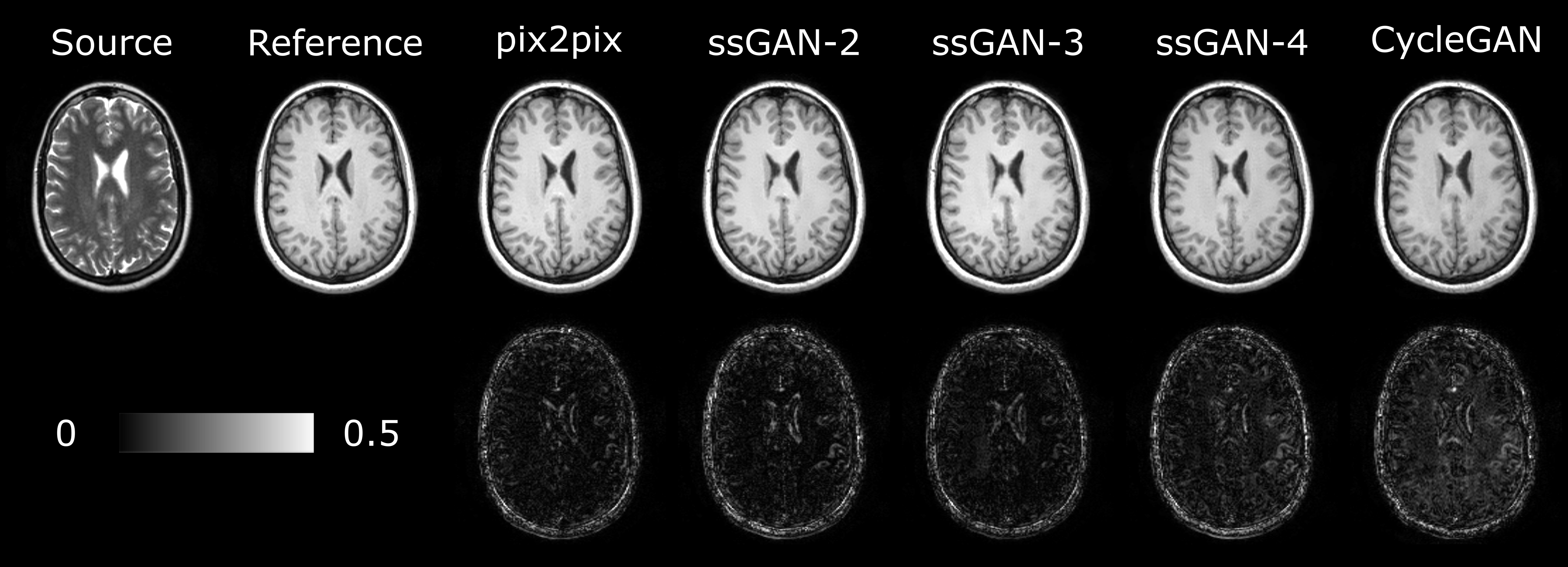}}
\caption{ssGAN was demonstrated on IXI for \TtwoTone~mapping against pix2pix and CycleGAN with ($R_{source}=1$). Synthesized images from ssGAN-2, ssGAN-3, ssGAN-4, pix2pix, and CycleGAN are displayed together with the reference and source images in the first row. The corresponding error maps for the synthesized images are displayed in the second row.}
\label{fig:comp_cycpix_supp}
\end{figure*}

\clearpage
\newpage
\begin{figure*}[htp]
\centerline{\includegraphics[width=0.99\textwidth]{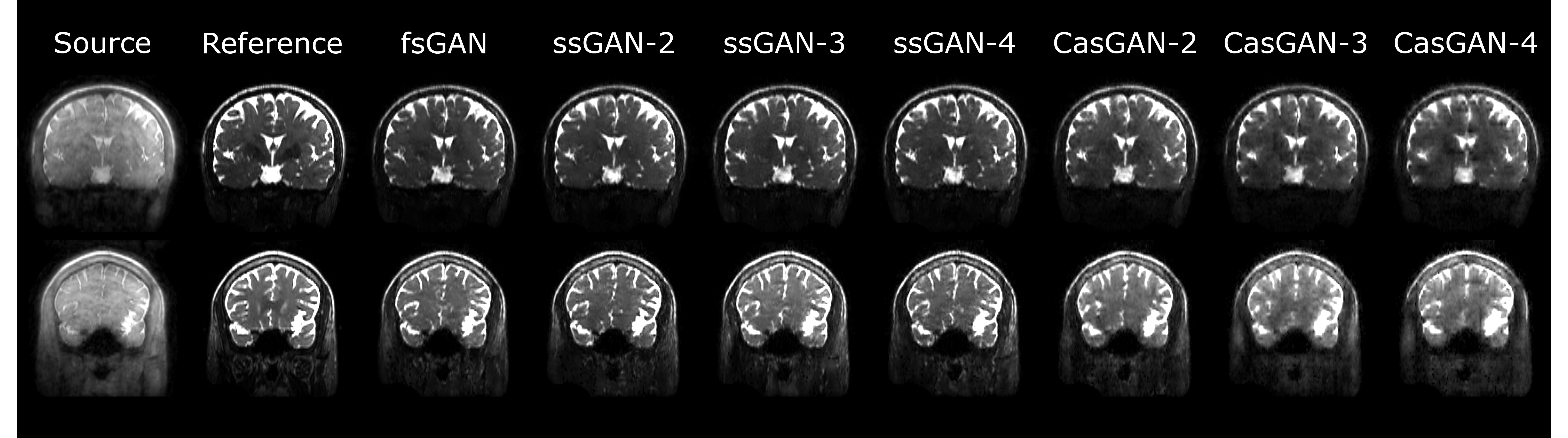}}
\caption{The proposed ssGAN model was demonstrated on the in vivo brain dataset for multi-coil complex \PDTtwo~synthesis task with source contrast acquisitions undersampled by $R_{target}=2$. Representative results from two different subjects are shown in separate rows. Synthesized coil-combined images from fsGAN, ssGAN, and CasGAN are displayed along with the source image and the reference image.}
\label{fig}
\end{figure*}

\clearpage
\newpage
\renewcommand{\tabcolsep}{3pt}
\renewcommand{\arraystretch}{1.40}
\begin{table*}[h]
\begin{center}
\centering
\caption{Effects of Tensor Losses on Synthesis Quality}
\begin{tabular}{ccccccc}
\cline{2-7}
                    & \multicolumn{3}{c}{T1 to T2} & \multicolumn{3}{c}{T2 to T1} \\ \cline{2-7} 
                    & PSNR     & SSIM    & MSE     & PSNR     & SSIM    & MSE     \\ \hline
ssGAN               & 25.25    & 90.75   & 0.32    & 26.47    & 92.27   & 0.24    \\
ssGAN (w/o image)   & 25.12    & 90.36   & 0.328   & 26.12    & 91.78   & 0.264   \\
ssGAN (w/o k-space) & 24.96    & 90.30   & 0.341   & 26.41    & 92.19   & 0.246   \\ \hline
                    & \multicolumn{3}{c}{T1 to T2} & \multicolumn{3}{c}{T2 to T1} \\ \cline{2-7} 
                    & \multicolumn{3}{c}{FID}      & \multicolumn{3}{c}{FID}      \\ \hline
ssGAN               & \multicolumn{3}{c}{22.89}    & \multicolumn{3}{c}{16.80}    \\
ssGAN (w/o adv)     & \multicolumn{3}{c}{23.13}         & \multicolumn{3}{c}{24.35}    \\ \hline
\end{tabular}
\end{center}
\end{table*}

\end{document}